\documentclass{article}
\usepackage[utf8]{inputenc}
\usepackage{url}

\title{The Myth of Complete AI-Fairness}
\author{Virginia Dignum}
\date{Ume{\aa} University, Sweden\\Email: \textit{virginia@cs.umu.se}}

\begin{document}

\maketitle

Just recently, IBM invited me to participate in a panel titled \textit{``Will AI ever be completely fair?”} My first reaction was that it surely would be a very short panel, as the only possible answer is `no'. In this short paper, I wish to further motivate my position in that debate: ``I will never be completely fair. Nothing ever is. The point is not complete fairness, but the need to establish metrics and thresholds for fairness that ensure trust in AI systems".

The idea of fairness and justice has long and deep roots in Western civilization, and is strongly linked to ethics. It is therefore not strange that it is core to the current discussion about the ethics of development and use of AI systems. Given that we often associate fairness with consistency and accuracy, the idea that our decisions and decisions affecting us can become fairer by replacing human judgement by automated, numerical, systems, is therefore appealing. However, as Laurie Anderson recently said\footnote{As quoted by Kate Crawford on Twitter \url{https://twitter.com/katecrawford/status/1377551240146522115}; 1 April 2021.} ``If you think technology will solve your problems, you don’t understand technology — and you don’t understand your problems."  AI is not magic, and its results are fundamentally constrained by the convictions and expectations of those that build, manage, deploy and use it. Which makes it crucial that we understand the mechanisms behind the systems and their decision capabilities.

The pursuit of fair AI is currently a lively one. One involving many researchers, meetings and conferences (of which FAccT\footnote{\url{https://facctconference.org/}} is the most known) and refers to the notion that an algorithm is fair, if its results are independent of given variables, especially those considered sensitive, such as the traits of individuals which should not correlate with the outcome (i.e. gender, ethnicity, sexual orientation, disability, etc.). However, nothing is ever 100\% fair in 100\% of the situations, and due to complex networked connection, to ensure fairness for one (group) may lead to unfairness for others. Moreover, what we consider fair often does depend on the traits of individuals. An obvious example are social services. Most people believe in the need for some form of social services, whether it is for children, for the elderly, for the sick or the poor. And many of us will benefit from social services at least once in our lives. Decision making in the attribution of social benefits is dependent on individual characteristics such as age, income, or chronic health problems. Algorithmic fairness approaches however overemphasize concepts such as equality and do not adequately address caring and concern for others.

Many years ago, I participated in a project at my children basic school that was aimed at helping kids develop fairness standards, roughly modelled along Kohlberg’s stages of moral development. It became clear quite quickly that children aged 6-12 easily understand that fairness comes in many ‘flavours’: if given cookies to divide between all kids of the class, the leading principle was equality, i.e. giving each kid the same amount of cookies. But they also understood and accepted the concept of equity: for instance in deciding that a schoolmate with dyslexia should be given more time to perform a school exam. Unfortunately, for the average algorithm, common sense and world knowledge is many light years away from that of a six year old, and switching between equity and equality depending on what is the best approach to fairness in a given situation, is rarely a feature of algorithmic decision making. 

\section*{So, how does fairness work in algorithms and what is being done to correct for unfair results?}

Doctors deciding on a patient’s treatment, or judges deciding on sentencing, must be certain that probability estimates for different conditions are correct for each specific subject, independent of age, race or gender. Increasingly these decisions are mediated by algorithms. Algorithmic fairness can be informally described as the probability of being classified in a certain category should be similar to for all that exhibit those characteristics, independently of other traits or properties. In order to ensure algorithmic fairness, given often very unbalanced datasets, data scientists use calibration (i.e. the comparison of the actual output and the expected output). Moreover, if we are concerned about fairness between two groups (e.g. male and female patients, or African-American defendants and Caucasian defendants) then this calibration condition should hold simultaneously for the set of people within each of these groups as well~\cite{flores2016false}.
Calibration is a crucial condition for risk prediction tools in many settings. If a risk prediction tool for evaluating defendants is not calibrated towards race, for example, then a probability estimate could carry different meaning for African-American than for Caucasian defendants, and hence the tool would have the unintended and highly undesirable consequence of incentivizing judges to take race into account when interpreting the tool’s predictions~\cite{pleiss2017fairness}. At the same time, ideally the incidence of false positives (being incorrectly classified as ‘X’) and false negatives (failing incorrectly to be classified as ‘X’) should be the same independently of other traits or properties. That is, fairness also means that, for instance, male and female candidates have the same chance of being offered a, for them irrelevant, service or product, or failing to receive for them relevant services or products.

Unfortunately, research shows that it is not possible to satisfy some of these expected properties of fairness simultaneously: calibration between groups, balance for false negatives, and balance for false positives. This means that if we calibrate data, we need to be prepared to accept higher levels of false positives and false negatives for some groups, and to deal with their human and societal impact~\cite{kleinberg2016inherent}. Taking the diagnostic example, a false positive means that a patient is diagnosed with a disease they don’t have. With a false negative that disease goes undiagnosed. The impact of either, both personal as well as societal, can be huge. In the same way, being wrongly classified as someone with a high risk to re-offend(false positive) has profound personal consequences, given that innocent people are held without bail, while incorrect classification as someone with a low risk to re-offend (false negative) has deep societal consequences, where people that pose a real criminal threat are let free\footnote{This example is at the core of the well-known Propublica investigations of the COMPAS algorithms used by courts in the US to determine recidivism risk: \url{www.propublica.org/article/how-we-analyzed-the-compasrecidivism-algorithm}.}. 

Given these technical difficulties in achieving perfectly fair, data-driven, algorithms, it is high time to start a conversation about the societal and individual impact of false positives and false negatives, and, more importantly, about what should be the threshold for acceptation of algorithmic decisions, that, by their nature, will never be completely ‘fair’.

\section*{Fairness is not about bias but about prejudice}

A commonly voiced explanation for algorithmic bias is the prevalence of human bias in the data. For example, when a job application filtering tool is trained on decisions made by humans, the machine learning algorithm may learn to discriminate against women or individuals with a certain ethnic background. Often this will happen even if ethnicity or gender are excluded from the data since the algorithm will be able to exploit the information in the applicant’s name, address or even the use of certain words. For example, Amazon’s recruiting AI system filtered out applications by women, because they lacked ‘masculine’ wording, commonly used in applications by men.

There are many reasons for bias in datasets, from choice of subjects, to the omission of certain characteristics or variables that properly capture the phenomenon we want to predict, to changes over time, place or situation, to the way training data is selected. 

Much has been done already to categorize and address the many forms of machine bias~\cite{schnabel2016recommendations}. Also, many tools are available to support to unbias AI systems, including IBM’s AI Fairness 360\footnote{\url{https://github.com/Trusted-AI/AIF360}} and Google’s What If Tool\footnote{\url{https://pair-code.github.io/what-if-tool/index.html#about}}. Basically, these tools support the testing and mitigation of bias through libraries of methods and test environments. According to  Google “[...] with the What If Tool you can test performance in hypothetical situations, analyse the importance of different data features, and visualize model behaviour across multiple models and subsets of input data, and for different ML fairness metrics.” Note the focus on performance, a constant in much of the work on AI.

However, not all bias is bad, in fact, there are even biases in the way we approach bias. Bias in human data in not only impossible to fully eliminate, it is often there for a reason. Bias is part of our lives partly because, we do not have enough cognitive bandwidth to make every decision from ground zero and therefore need to use generalizations, or biases, as a starting point. Without bias, we would not been able to survive as a species, it helps us selecting from a myriad of options in our environment. But not all biases are bad. But that doesn’t mean we shouldn’t address them.

Bias is not the problem, prejudice and discrimination are. Whereas prejudice represents a preconceived judgment or attitude, discrimination is a behaviour. In society, discrimination is often enacted through institutional structures and policies, and embedded in cultural beliefs and representations, and is thus reflected in any data collected. The focus need be on using AI to support interventions aimed at reducing prejudice and discrimination, e.g. through education, facilitation of intergroup contact, targeting social norms promoting positive relations between groups, or supporting people identify their own bias and prejudices.

Facial analysis tools and recognition software have raised concerns about racial bias in the technology. Work by Joy Buolamwini and Timnit Gebru has shown how deep these biases go and how hard they are to eliminate~\cite{buolamwini2018gender}. In fact, debiasing AI often leads to other biases. Sometimes this is known and understood, such as the dataset they created as alternative to the datasets commonly used for training facial recognition algorithms: using what they called 'parliaments', Buolamwini and Gebru created a dataset of faces balanced in terms of race and gender, but notably unbalanced in terms of age, lighting or pose. As long as this is understood, this dataset is probably perfectly usable for training an algorithm to recognise faces of a certain age, displayed under the same lighting conditions and with the same pose. It will however not be usable if someone tries to train an algorithm to recognise children’s faces. This illustrates that debiasing data is not without risks, in particular because it focus on those characteristics that we are aware of, which are ‘coloured’ by our own experience, time, place and culture. 

Moreover, AI bias is more than biased data. It starts with who is collecting the data, who is involved in selecting and/or designing the algorithms and who is training the algorithms and labelling the data. Moreover, decisions about which and whose data is collected and which data is being used to train the algorithms and how the algorithm is designed and trained also influence the fairness of the results. 
From labelling farms to ghost workers, the legion of poorly paid, badly treated and ignored human labourers, working behind the scenes of any working AI system, is huge and little is being done to acknowledge them and to improve their working conditions. Books such as ‘Ghost work’~\cite{gray2019ghost}  by Mary L. Gray and Siddharth Suri, or ‘Atlas of AI’~\cite{crawford2021atlas}  by Kate Crawford, are raising the issue but, as often is the case, the ‘folk is sleeping’: it is easier to use the systems and profit from their results, than to question how these results are being achieved and at what cost. The question is thus: how fair is algorithm fairness for those that label, train and calibrate the data it needs to produce fair results and, more importantly, if we expect them to provide us with unbiased data, shouldn’t we be treating them fairly?

\section*{Beyond fairness}

The fact that algorithms and humans, cannot ever be completely fair, does not mean that we should’ve up and accept it. Improving fairness and overcoming prejudice is partly a matter of understanding how the technology works. A matter of education. Moreover, using technology properly, fair treatment of those using and being affected by it, requires participation. Still many stakeholder are not invited to the table, not joining the conversation.

Lack of fairness in AI systems is often also linked to a lack of explanatory capabilities. If the results of the system cannot be easily understood or explained, it is difficult to assess its fairness. Many of the current tools that evaluate bias and fairness help identify where biases may occur, whether in the data or the algorithms or even in their testing and evaluation. Even if not all AI systems can be fully explainable, it is important to make sure that their decisions are reproducible and the conditions for their use are clear and open to auditing.  

Current AI algorithms are built for accuracy and performance, or for efficiency. Improving the speed of the algorithm, minimizing its computational requirements and maximizing the accuracy of the results are the mantras that lead current computer science and engineering education. However, these are not the only optimization criteria. When humans and society are at stake, other criteria need be considered. How do you balance safety and privacy? Explainability and energy resources? Autonomy and accuracy? What do you do when you cannot have both? Such moral overload dilemmas are at the core of responsible development and use of AI~\cite{dignum2019responsible}. 

Addressing them requires multidisciplinary development teams and involvement of the humanities and social sciences in software engineering education. It also requires a redefinition of incentives and metrics for what is a ‘good’ system. Doing the right thing, and doing it well means that we also need to define what is good and for whom.

Finally, it is important to keep continuous efforts to improve algorithms and data, define regulation and standardisation, and develop evaluation tools and corrective frameworks. But the same time, we cannot ignore that no technology is without risk, no action is without risk. It is high time to start the conversation on which AI-risks we find acceptable for individuals and for society as a whole, and how we distribute these risks, as well as the benefits of AI.

\bibliographystyle{plain}
\bibliography{refs}

\end{document}